\pdfoutput=1
\documentclass[12pt]{article}

\font\grande=cmr9.5 scaled \magstep4
\font\medio=cmr9.5 scaled \magstep2
\outer\def\beginsection#1\par{\medbreak\bigskip
      \message{#1}\leftline{\bf#1}\nobreak\medskip
\vskip-\parskip
      \noindent}
\usepackage{graphicx} 
\begin{document}
\bibliographystyle {unsrt}

\titlepage

\begin{flushright}
\end{flushright}

\begin{center}
{\grande Effective horizons, junction conditions}\\
\vspace{5mm}
{\grande and large-scale magnetism}\\
\vspace{1.5cm}
 Massimo Giovannini
 \footnote{Electronic address: massimo.giovannini@cern.ch}\\
\vspace{1cm}
{{\sl Department of Physics,  
Theory Division, CERN, 1211 Geneva 23, Switzerland }}\\
\vspace{0.5cm}
{{\sl INFN, Section of Milan-Bicocca, 20126 Milan, Italy}}
\vspace*{0.5cm}
\end{center}

\vskip 0.5cm
\centerline{\medio  Abstract}
The quantum mechanical generation of hypermagnetic and hyperlectric fields
in four-dimensional conformally flat background geometries 
rests on the simultaneous continuity of the effective horizon and 
of the extrinsic curvature across the inflationary boundary. 
The junction conditions for the gauge fields are derived in general 
terms and corroborated by explicit examples with particular attention to the 
limit of a sudden (but nonetheless continuous) transition of the effective horizon. 
After reducing the dynamics to a pair of integral equations related
by duality transformations, we compute the power spectra and deduce a 
novel class of logarithmic corrections which turn out to be, however, 
numerically insignificant and overwhelmed by the conductivity effects once the gauge 
modes reenter the effective horizon. In this perspective the magnetogenesis requirements 
and the role of the postinflationary conductivity are clarified and reappraised. 
As long as the total duration of the inflationary phase is nearly minimal, 
quasi-flat hypermagnetic power spectra are comparatively 
more common than in the case of vacuum initial data.
\vskip 0.5cm

\noindent

\vspace{5mm}

\vfill
\newpage
The qualitative description of large-scale cosmological perturbations \cite{wein1,wein2,primer} 
stipulates that a given wavelength exits the Hubble radius at some typical conformal time 
$\tau_{ex}$ during an inflationary stage of expansion and approximately reenters 
at $\tau_{re}$, when the Universe still expands but in a decelerated manner. 
By a mode being beyond the horizon we only mean that the physical wavenumber 
is much less than the expansion rate:  this does not necessarily 
have anything to do with causality \cite{wein2}. Indeed, the initial conditions of the 
Einstein-Boltzmann hierarchy (mandatory for the calculation 
of the temperature and polarization anisotropies) are set when the relevant modes 
are larger than the Hubble radius prior to matter-radiation equality \cite{primer}. 
Similarly the physical wavenumbers of the hyperelectric and hypermagnetic fields 
can be much smaller than the rate of variation of the susceptibility ($\chi$ in what follows) 
which now plays the role of the effective horizon. The junction conditions  
of the gauge power spectra will be derived in general terms and then corroborated by 
specific examples with particular attention to the the case of sudden 
(but continuous) postinflationary transitions. Using the obtained results the gauge power 
spectra will be computed in the case of generalized quantum mechanical initial conditions 
of the hypercharge field.

The four-dimensional action discussed in \cite{action1} concisely summarizes a 
large class of magnetogenesis scenarios and it can be written, for the present ends, 
as\footnote{We shall be working in a conformally flat background geometry 
$\overline{g}_{\mu\nu} = a^2(\tau)\eta_{\mu\nu}$ where $\eta_{\mu\nu}$ denotes the Minkowski metric, 
$a(\tau)$ is the scale factor and $\tau$ parametrizes the conformal time coordinate. 
The components of the Abelian field strength appearing in Eq. (\ref{first}) are $Y^{0i} = e^{i}/a^2$ and 
$Y^{ij}= -\epsilon^{ijk} b_{k}/a^2$.  The canonical electric and magnetic fields of Eq. (\ref{second}) are defined as 
$\vec{B} = a^2 \, \sqrt{\lambda}\, \vec{b}$ and  $\vec{E} = a^2 \, \sqrt{\lambda}\, \vec{e}$.}:
\begin{equation}
S = - \frac{1}{16 \pi} \int \, d^{4} x\, \sqrt{-g}\biggl[ {\mathcal M}_{\sigma}^{\rho} Y_{\alpha\rho} \, Y^{\alpha\sigma}  - 
{\mathcal N}_{\sigma}^{\rho} \widetilde{Y}_{\alpha\rho} \, \widetilde{Y}^{\alpha\sigma}  \biggr],
\label{first}
\end{equation}
where $g$ denotes the determinant of the four-dimensional metric\footnote{We shall be working in a conformally flat background geometry 
$\overline{g}_{\mu\nu} = a^2(\tau)\eta_{\mu\nu}$ where $\eta_{\mu\nu}$ denotes the Minkowski metric, 
$a(\tau)$ is the scale factor and $\tau$ parametrizes the conformal time coordinate. 
The components of the Abelian field strength appearing in Eq. (\ref{first}) are $Y^{0i} = e^{i}/a^2$ and 
$Y^{ij}= -\epsilon^{ijk} b_{k}/a^2$.  The canonical electric and magnetic fields of Eq. (\ref{second}) are defined as 
$\vec{B} = a^2 \, \sqrt{\lambda}\, \vec{b}$ and  $\vec{E} = a^2 \, \sqrt{\lambda}\, \vec{e}$.}; 
$Y_{\alpha\beta}$  and $\widetilde{Y}^{\alpha\beta}$ are, respectively, the gauge field strength and its dual.
While the two symmetric tensors ${\mathcal M}_{\sigma}^{\rho}$ and ${\mathcal N}_{\sigma}^{\rho}$ parametrize, 
in full generality, the dependence upon the electric and magnetic susceptibilities, Eq. (\ref{first}) includes, 
as a special case, the derivative couplings typical of the relativistic theory of Casimir-Polder and 
Van der Waals interactions \cite{such}. Even though the whole discussion could be carried on 
in the case of unequal magnetic and electric susceptibilities by using the results reported in \cite{action1},  for 
the sake of simplicity the attention will now be focussed on the case 
${\mathcal M}_{\sigma}^{\rho} = {\mathcal N}_{\sigma}^{\rho}  = (\lambda/2) \delta_{\sigma}^{\rho}$. 
In this instance the evolution equations derived from Eq. (\ref{first}) are:
\begin{equation}
 \vec{E}^{\,\prime} +  {\mathcal F} \vec{E} = \vec{\nabla}\times \vec{B}, \qquad 
 \vec{B}^{\,\prime} - {\mathcal F} \vec{B} =  - \vec{\nabla}\times \vec{E},\qquad 
 {\mathcal F} = \frac{\chi^{\,\prime}}{\chi}, 
\label{second}
\end{equation}
where, as already mentioned, $\chi =  \sqrt{\lambda}$ represents the susceptibility and the 
prime denotes a derivation with respect to the conformal time coordinate. 
As implied by the duality symmetry \cite{duality1}, when $\chi\to 1/\chi$ 
(i.e. ${\mathcal F} \to - {\mathcal F}$) the two equations appearing in 
Eq. (\ref{second}) are interchanged  provided $\vec{E} \to - \vec{B}$ and  $\vec{B} \to \vec{E}$.  
Equations (\ref{first}) and (\ref{second}) contain, as a particular case, a class of magnetogenesis 
models based on the evolution of the inflaton or of some other spectator field (see, e. g. \cite{DT1,DT2,DT3} 
for an incomplete list of references\footnote{Equations (\ref{first}) and (\ref{second}) do not include the interesting case of a spectator 
Higgs field non-minimally coupled 
to gravity and possibly leading to sizable magnetic fields ${\mathcal O}(10^{-20})$ 
G for the benchmark scale of the protogalactic collapse \cite{DT4}.}). Various scenarios 
aim at producing magnetic fields with approximate intensities 
of a few hundredths of a nG ($1 \, \mathrm{nG} = 10^{-9}\, \mathrm{G}$) and over typical 
comoving scales between few Mpc and 100 Mpc.
When the intensities are much lower than ${\mathcal O}(10^{-3})$ nG a dynamo action (of some sort) seems 
mandatory (see, for instance, Ref. \cite{rev} for a time ordered but still incomplete list of review articles).

In conformally flat backgrounds geometries of Friedmann-Robertson-Walker type
 the Coulomb gauge condition (i.e. $Y_{0} =0$  and $\vec{\nabla}\cdot \vec{Y} =0$)  is preferable since it 
 is preserved (unlike the Lorentz gauge condition) by a conformal rescaling of the metric; with this choice,
 when the electric and magnetic susceptibilities coincide, Eq. (\ref{first}) reduces to:
\begin{equation}
S = \frac{1}{2}\int \,d\tau\, d^{3} x \, \biggl\{ \vec{{\mathcal A}}^{\,\prime \,2} + {\mathcal F}^2 
 \vec{{\mathcal A}}^{\,2}  - 2 {\mathcal F} \vec{{\mathcal A}} \cdot \vec{{\mathcal A}}^{\,\prime} -
  \partial_{i} \vec{{\mathcal A}} \cdot \partial^{i} \vec{{\mathcal A}}\biggr\},
\label{third}
\end{equation}
where $\vec{{\mathcal A}} = \sqrt{ \lambda/(4\pi)} \vec{Y}$. In terms of $\vec{{\mathcal A}}$ and of its conjugate 
momentum $\vec{\Pi}$ the 
canonical Hamiltonian derived from the action (\ref{third}) is:
\begin{equation}
H(\tau) = \frac{1}{2} \int d^3 x \biggl[ \vec{\Pi}^{2} + 2 \,{\mathcal F}\, \vec{\Pi} \cdot \vec{{\mathcal A}} + 
 \partial_{i} \vec{{\mathcal A}} \cdot \partial^{i} \vec{{\mathcal A}}\biggr], 
 \qquad \vec{\Pi} = \vec{{\mathcal A}}^{\,\prime} - {\mathcal F} \vec{{\mathcal A}}.
\label{fourth}
\end{equation}
In terms of the normal modes $\vec{{\mathcal A}}$ he hyperelectric and hypermagnetic fields of Eq. (\ref{second}) are defined, respectively, as  $\vec{E}= - \vec{\Pi}(\tau,\vec{x})$ and 
$\vec{B} = \vec{\nabla}\times \vec{{\mathcal A}}$; the corresponding field operators in the Heisenberg 
description are:
\begin{eqnarray}
&& \hat{B}_{i}(\tau, \vec{x}) = - \frac{i\, \epsilon_{m n i}}{(2\pi)^{3/2}}  \sum_{\alpha} \int d^{3} k \,k_{m} \,e^{(\alpha)}_{n} \,
\biggl[ f_{k}(\tau)\, \hat{a}_{\vec{k}, \alpha} e^{- i \vec{k} \cdot \vec{x}}  - f_{k}^{*}(\tau) \hat{a}^{\dagger}_{\vec{k}, \alpha}
e^{ i \vec{k} \cdot \vec{x}}\biggr], 
\label{rel2}\\
&& \hat{E}_{i}(\tau,\vec{x}) = - \frac{1}{(2\pi)^{3/2}}  \sum_{\alpha} \int d^{3} k \,e^{(\alpha)}_{i} \,
\biggl[ g_{k}(\tau)  \hat{a}_{\vec{k}, \alpha} e^{- i \vec{k} \cdot \vec{x}}  + g_{k}^{*}(\tau) \hat{a}^{\dagger}_{\vec{k}, \alpha}e^{ i \vec{k} \cdot \vec{x}} \biggr],
\label{rel3}
\end{eqnarray}
where the sum is performed over the physical polarizations $e^{(\alpha)}_{i}$ while the mode functions 
$f_{k}$ and $g_{k}$ obey, in the absence of conductivity, the following pair of dual equations:
\begin{equation}
f_{k}' = g_{k} + {\mathcal F} f_{k},\qquad  g_{k}' = -  k^2 f_{k} - {\mathcal F} g_{k}.
\label{oldeq}
\end{equation}
From Eq. (\ref{oldeq})  two (decoupled) second-order differential equations can be derived for $f_{k}$ (i.e. 
$f_{k}^{\prime\prime} + [ k^2 - \chi^{\prime\prime}/\chi] f_{k} =0$) and  for $g_{k}$ (i.e. $g_{k}^{\prime\prime} 
+ [ k^2 - \chi (1/\chi)^{\prime\prime}] g_{k} =0$).  To guarantee the correct formulation of the Cauchy problem
the initial conditions must be assigned in agreement with $g_{k} = f_{k}^{\prime} - {\mathcal F} f_{k}$  but without
 the continuity of  $\chi$ (and of its first derivative) the pump fields $\chi^{\prime\prime}/\chi$ and $ \chi (\chi^{-1})^{\prime\prime}$
will be singular at the transition points. Moreover,  to enforce the canonical form of the 
commutation relations\footnote{ The (equal time) commutation relations (in units $\hbar = c =1$) read
$[\hat{{\mathcal A}}_{i}(\vec{x}_{1},\tau),\hat{\Pi}_{j}(\vec{x}_{2},\tau)] = i \Delta_{ij}(\vec{x}_{1} - \vec{x}_{2})$.
Defining, as usual, $P_{ij}(k) = (\delta_{ij} - k_{i} k_{j}/k^2)$ the function $\Delta_{ij}(\vec{x}_{1} - \vec{x}_{2}) = 
\int d^{3}k e^{i \vec{k} \cdot (\vec{x}_{1} - \vec{x}_2)} P_{ij}(k)/ (2\pi)^3$ is the transverse generalization of the Dirac delta function. }
the Wronskian ${\mathcal W}(\tau) = f_{k}(\tau)g_{k}^{*}(\tau) - f_{k}^{*}(\tau) g_{k}(\tau)$ (conserved and invariant under duality) 
must be normalized as ${\mathcal W}(\tau) =i$. 

Using Eqs. (\ref{rel2}) and (\ref{rel3}) the magnetic and electric field operators 
are expressible in Fourier space and their expectation values at coincident conformal times are:
\begin{eqnarray}
&& \langle \hat{B}_{i}(\tau,\vec{k})\, \hat{B}_{j}(\tau,\vec{p}) \rangle = \frac{2\pi^2}{k^3}\, 
P_{B}(k,\tau)\, P_{ij}(k)  \,\delta^{(3)}(\vec{k} + \vec{p}),\qquad P_{B}(k,\tau) = \frac{k^{5}}{2\, \pi^2\,} \, |f_{k}(\tau)|^2, 
\label{cc1}\\
&& \langle \hat{E}_{i}(\tau,\vec{k})\, \hat{E}_{j}(\tau,\vec{p}) \rangle = \frac{2\pi^2}{k^3}\, 
P_{E}(k,\tau)\, P_{ij}(k)  \,\delta^{(3)}(\vec{k} + \vec{p}),\qquad P_{E}(k,\tau) = \frac{k^3}{2\, \pi^2\,} \,  |g_{k}(\tau)|^2,
\label{cc2}
\end{eqnarray}
where $P_{B}(k,\tau)$ and $P_{E}(k,\tau)$ denote respectively the hypermagnetic and the hyperelectric power 
spectra\footnote{As in the case of Eq. (\ref{second}), when ${\mathcal F} \to - {\mathcal F}$ (i.e. $\chi\to 1/\chi$), the equations of Eq. (\ref{oldeq}) 
  are interchanged provided  $f_{k} \to g_{k}/k$ and $g_{k} \to - k f_{k}$ (see Eq. (\ref{oldeq})). 
Under the same duality transformation Eqs. (\ref{cc1})--(\ref{cc2}) imply that $P_{B}\to P_{E}$ and vice versa. Again this
symmetry \cite{duality1} is verified provided $\chi$ and $\chi^{\prime}$ are simultaneously continuous everywhere and, in particular, 
across the inflationary boundary.} that shall now be derived without relying  
on the nature of the transition but only on the overall continuity and differentiability of the evolution. For this purpose Eq. (\ref{oldeq}) can be 
transformed into a pair of integral equations with initial conditions assigned at $\tau_{ex}$:
\begin{eqnarray}
f_{k}(\tau) &=& \frac{\chi(\tau)}{\chi_{ex}} \biggl\{ f_{k}(\tau_{ex}) 
+ \biggl[ f_{k}^{\prime}(\tau_{ex}) - {\mathcal F}_{ex} f_{k}(\tau_{ex})\biggr] \int_{-\tau_{ex}}^{\tau} \frac{\chi_{ex}^2}{\chi^2(\tau_{1})} d\tau_{1}
\nonumber\\
&-&k^2 \int_{-\tau_{ex}}^{\tau} \frac{d\tau_{1}}{\chi^2(\tau_{1})} \int_{-\tau_{ex}}^{\tau_{1}} \chi_{ex} \chi(\tau_{2}) f_{k}(\tau_{2}) d\tau_{2} \biggr\},
\label{INT1}\\
g_{k}(\tau) &=& \frac{\chi_{ex}}{\chi(\tau)} \biggl\{ g_{k}(\tau_{ex}) 
+ \biggl[ g_{k}^{\prime}(\tau_{ex}) + {\mathcal F}_{ex} g_{k}(\tau_{ex})\biggr] \int_{-\tau_{ex}}^{\tau} \frac{\chi^2(\tau_{1})}{\chi_{ex}^2} d\tau_{1}
\nonumber\\
&-&k^2 \int_{-\tau_{ex}}^{\tau} d\tau_{1}\chi^2(\tau_{1}) \int_{-\tau_{ex}}^{\tau_{1}}  \frac{g_{k}(\tau_{2})}{\chi_{ex} \chi(\tau_{2})}d\tau_{2} \biggr\}.
\label{INT2}
\end{eqnarray}
Depending on the evolution of the background (either before or after $\tau_{i}$) the 
condition $k^2 = |\chi^{\prime\prime}/\chi|$ defines either $\tau_{ex}$ or $\tau_{re}$; 
the latter condition can also be dubbed, after simple algebra, as:
\begin{equation}
k^2 =a^2 F^2 \biggl( 1 + \frac{H}{F} - \epsilon_{F}\biggr), \qquad \epsilon_{F} = - \frac{\dot{F}}{F^2},
\label{turning}
\end{equation}
where the overdot denotes a derivation with respect to the cosmic time coordinate; moreover
$H= \dot{a}/a = {\mathcal H}/a$ is the Hubble rate while 
$F = \dot{\chi}/\chi={\mathcal F}/a$ is the rate of variation of $\chi$.
In Eq. (\ref{turning}) $\epsilon_{F}$ is the analog of the 
conventional slow-roll parameter (i.e. $ \epsilon_{H} = - \dot{H}/H^2$).

Inside the effective horizon (i.e. $k /{\mathcal F} \ll 1$)  the initial conditions for $\tau \leq - \tau_{ex}$ appearing in Eqs. (\ref{INT1}) and (\ref{INT2}) 
are plane waves (i.e. $f_{k}(\tau) = [ b_{+}(k) e^{- i k\tau} + b_{-}(k) e^{i k\tau}]/\sqrt{2 k}$)
solving Eq. (\ref{oldeq}) for $k\ll {\mathcal F}$. The vacuum Cauchy data correspond to  $b_{+}(k) \to 1$  
and $b_{-}(k) \to 0$. Conversely when $b_{+}(k) \neq 1$ and $b_{-}(k)\neq 0$ the mode functions for $\tau < \tau_{ex}$
correspond to an initial state whose average multiplicity is $|b_{-}(k)|^2$.
The iterative solution of Eqs. (\ref{INT1}) and (\ref{INT2}) can then be obtained to the wanted order in $k^2 \tau^2$ 
but the lowest order solution reduces to the evaluation of the following pair of (dual) integrals:
\begin{equation}
{\mathcal I}_{f}(\tau_{ex}, \tau_{re}) = \int_{-\tau_{ex}}^{\tau_{re}} \frac{\chi_{ex}^2}{\chi^2(\tau)} d\tau, 
\qquad {\mathcal I}_{g}(\tau_{ex}, \tau_{re}) 
= \int_{-\tau_{ex}}^{\tau_{re}} \frac{\chi^2(\tau)}{\chi_{ex}^2} d\tau,
\label{FG}
\end{equation}
which are both defined provided the integrand is (at least) continuous. 
Because of this property the evolution across the inflationary boundary can be 
globally described by introducing the following averages of $\epsilon_{F}$ and of $H/F$ namely
\begin{equation}
 \overline{\epsilon}_{F}  = \int_{-\tau_{ex}}^{\tau_{re}} \biggl(- \frac{\dot{F}}{F^2}\biggr) 
 \frac{ d \tau}{\chi^2}\, \biggl/ \,\int_{\tau_{ex}}^{\tau_{re}}  \frac{ d \tau}{\chi^2},\qquad 
 \langle \frac{H}{F} \rangle =  \int_{-\tau_{ex}}^{\tau_{re}} \biggl(\frac{H}{F}\biggr) 
 \frac{ d \tau}{\chi^2}\biggl/ \,\int_{\tau_{ex}}^{\tau_{re}} \frac{ d \tau}{\chi^2}.
 \label{int2c}
 \end{equation}
Recalling Eq. (\ref{int2c}) and that $\chi$ and ${\mathcal F}$ are continuous everywhere 
(and in particular across the inflationary  boundary),
after two integrations by parts the integral ${\mathcal I}_{f}(\tau_{ex}, \tau_{re})$ becomes:
\begin{equation} 
{\mathcal I}_{f}(\tau_{ex}, \tau_{re}) = \frac{1}{2} \biggl(\frac{1}{{\mathcal F}_{ex}}-  
\frac{\chi_{ex}^2}{ \chi^2_{re} {\mathcal F}_{re}}\biggr)+\frac{1}{2} \biggl(  \overline{\epsilon}_{F} - 
\langle \frac{H}{F} \rangle \biggr) \int_{-\tau_{ex}}^{\tau_{re}} \frac{\chi_{ex}^2}{\chi^2(\tau)} d\tau.
\label{int2a}
\end{equation}
According to Eq. (\ref{int2a}) 
$\chi$ cannot freeze instantaneously to a constant value after the end of inflation, 
as observed in explicit numerical integrations (see, for instance, the last paper of Ref. \cite{DT2}).  
We now observe that ${\mathcal I}_{f}(\tau_{ex}, \tau_{re})$ multiplies the second term at the right hand side of 
Eq. (\ref{int2a}). As a consequence the wanted integral (appearing both at the right and at the left hand side of Eq. (\ref{int2a})) 
can be solely expressed in terms of $\overline{\epsilon}_{F}$ and  $\langle (H/F)\rangle$ and $f_{k}(\tau)$ is explicitly given by: 
\begin{equation}
f_{k}(\tau) = \frac{\chi(\tau)}{\chi_{ex}} \biggl[ f_{k}(\tau_{ex}) +   \frac{g_{k}(\tau_{ex})}{2 -  
\overline{\epsilon}_{F} + \langle (H/F) \rangle} \biggl(\frac{1}{{\mathcal F}_{ex}}-  \frac{\chi_{ex}^2}{ \chi^2_{re} {\mathcal F}_{re}}\biggr)\biggr] + {\mathcal O}(k^2 \tau^2).
\label{int2d}
\end{equation}
We shall now  parametrize the inflationary evolution of the susceptibility 
as\footnote{To avoid absolute values in the spectral slopes, we shall assume throughout that $\nu \geq 1$; this condition is anyway verified
in the illustrative examples discussed below.} 
 $\chi_{inf}(\tau) = \chi_{i} (- \tau/\tau_{i} )^{1/2 -\nu}$ for $\tau < -\tau_{i}$ (where $-\tau_{i}$ marks the end of 
 the inflationary phase). Conversely for 
$\tau\geq - \tau_{i}$ we shall just assume the continuity of $\chi$ and ${\mathcal F}$; with these premises
we obtain, quite generically\footnote{This is true, in particular, when the rate 
of the evolution of $\chi$ and the expansion rate of the background geometry are proportional to each other
(i.e. $F = \delta  H$). In this instance, $2 - \overline{\epsilon}_{F} - \langle (H/F) \rangle = 2 + (1- \overline{\epsilon}_{H})/\delta$
where now $\epsilon_{H} = - \dot{H}/H^2$ is the conventional slow-roll parameter already mentioned 
above. We are supposing here that  
the inflationary evolution (i.e. $\epsilon_{H} \ll1$) is replaced by a radiation-dominated phase (i.e.  $\epsilon_{H} \to 2$).} that $2 - \overline{\epsilon}_{F} + \langle (H/F) \rangle \neq 0$.  
All in all we can then say that the continuity properties of the transition imply that 
the hypermagnetic power spectra of Eqs. (\ref{cc1})--(\ref{cc2}) are given by:
\begin{equation}
P_{B}(k,\tau_{re}) \simeq H_{i}^4 a_{i}^4 |k\tau_{i}|^{5- 2 \nu}\biggl[ 1 + {\mathcal O}(|k\tau_{i}|^{2 \nu})\biggr],
\label{PSfirst}
\end{equation} 
where, for the sake of simplicity, the vacuum  initial conditions have been imposed by setting $|b_{+}(k)|=1$ and $|b_{-}(k)| =0$.
Except for specific values of $\delta$ (possibly leading to conspiratorial cancellations) Eq. (\ref{PSfirst}) 
determines the hypermagnetic power spectra up to overall factors of order $1$.  With similar considerations the electric power spectra can also be derived;
furthermore, since under duality $\chi\to 1/\chi$ and 
${\mathcal F} \to - {\mathcal F}$, we will also have that $\nu \to 1 - \nu$ implying, 
from Eq. (\ref{PSfirst}) and only using the symmetries of the problem, that $P_{E}(k,\tau_{re}) \propto |k\tau_{i}|^{7 - 2\nu}$.
Even if they will not be directly relevant for the present discussion we explicitly verified 
that the method described here correctly leads to the electric power spectra implied by the duality symmetry \cite{duality1}. 
Indeed, as soon as the gauge modes reenter the effective horizon duality is explicitly broken since 
the evolution equations will only contain electric (and not magnetic) sources \cite{incon}.

When the extrinsic curvature, the susceptibility and the effective horizon are (simultaneously and explicitly) 
continuous across the inflationary boundary, the general derivation leading to Eq. (\ref{PSfirst}) can be corroborated by specific examples.  For this purpose we shall the inflationary scale factor shall be expressed as $a_{inf}(\tau) = (- \tau/\tau_{i})^{- \gamma}$ for $\tau < - \tau_i$ (where $\gamma=1$ in the case of an exact de Sitter phase\footnote{During a quasi-de Sitter phase, 
the connection between the  conformal time coordinate and  the Hubble rate is given by ${\mathcal H} = aH = -1/[(1-\epsilon_{H})\tau]$ at least in the 
case when $\epsilon_{H}$ is constant.}). In the subsequent radiation epoch (i.e. for $\tau\geq -\tau_{i}$) the scale factor is  
given by $a_{rad}(\tau) = [\gamma \tau + (\gamma + 1)\tau_i]/\tau_i$. Since the scale factors and  their first time derivatives are explicitly continuous  [i.e.  $a_{rad}(-\tau_{i}) = a_{i}(-\tau_{i})$ and $a^{\prime}_{rad}(-\tau_{i}) = a^{\prime}_{inf}(- \tau_{i})$], the extrinsic curvature ${\mathcal H}/a$ is also continuous [i.e. ${\mathcal H}_{rad}(-\tau_{i}) = {\mathcal H}_{inf}(- \tau_{i})$]. One of the simplest situations compatible with a sudden transition stipulates that
the susceptibility approaches exponentially its (constant) asymptotic value; the explicit expressions of $\chi_{inf}(\tau)$ and $\chi_{rad}(\tau)$ are given, respectively, by:
\begin{eqnarray}
\chi_{inf}(\tau) &=& \chi_{i} \biggl(- \frac{\tau}{\tau_{i}} \biggr)^{1/2 -\nu}, \qquad \tau< - \tau_{i},
\label{A}\\
\chi_{rad}(\tau) &=& \chi_{i} \biggl[ C + D e^{- \beta(\tau/\tau_{i} +1)} \biggr], \qquad \tau \geq - \tau_{i},
\label{B}
\end{eqnarray}
where we defined, for the sake of conciseness, $C =[ 1 - (1-2 \nu)/(2\beta)]$ and $D= (1 - C)$.
Equations (\ref{A}) and (\ref{B}) imply the continuity of $\chi$ in $-\tau_{i}$ [i.e. $\chi_{inf}(-\tau_{i}) = \chi_{rad}(-\tau_{i})$] and of its 
first derivative [i.e. $\chi_{inf}^{\prime}(-\tau_{i}) = \chi_{rad}^{\prime}(-\tau_{i})$]. 
The constant value $\chi_{i}$ is approached at a rate controlled by $\beta$: when $\beta \leq 1$ the transition is delayed while for $\beta > 1$ the transition is sudden\footnote{Another natural choice would be a power-suppressed profile of the type $\chi_{rad}(\tau) = \chi_{i}[ \overline{C} + \overline{D} ( 1 + \tau/\tau_{i})^{-\alpha}]$ 
where $\overline{C} = [2(\alpha + \nu) -1]/(2 \alpha)$ and $\overline{D} = 1 - \overline{C}$. 
The parameter $\alpha\geq 1$  is, in this case, the analog of $\beta$.}. 
Thanks to Eqs. (\ref{A})--(\ref{B})  the dual integrals of Eq. (\ref{FG}) are
\begin{eqnarray}
{\mathcal I}_{f}(\tau_{ex}, \tau_{re}) &=& \tau_{ex} \biggl\{ \frac{1}{2 \nu} \biggl(1 - \frac{| k \tau_{i}|^{2\nu}}{q^{2 \nu}} \biggr) + 
\frac{| k \tau_{i}|^{2\nu}}{q^{2 \nu} \beta C^2 }\biggl[ \ln{( C e^{z_{re}} + D)} + D \biggl(\frac{1}{C e^{z_{re}} + D} - 1\biggr)\biggr]\biggr\},
\nonumber\\
{\mathcal I}_{g}(\tau_{ex}, \tau_{re}) &=& \tau_{ex} \biggl\{ \frac{1}{2 (1 - \nu)} \biggl(1 - \frac{| k \tau_{i}|^{2( 1 - \nu)}}{q^{2(1- \nu)}} \biggr) + 
\frac{ C^2 | k \tau_{i}|^{1- 2\nu}}{q^{2(1- \nu)} \beta}\biggl[ z_{re} + 2 \frac{D}{C}(1 - e^{ - z_{re}})  
\nonumber\\
&+& \frac{D^2}{2 C^2}(1 -  e^{- 2 z_{re}})
\biggr]\biggr\},
\label{int1cc}
\end{eqnarray}
where the shorthand notations $z_{re} = \beta (\tau_{re}/\tau_{i} +1)$ and   $k \tau_{ex} = q(\nu)=\sqrt{\nu^2 - 1/4}$ have been adopted.  
The second turning point  is determined by the analog of Eq. (\ref{turning}) (i.e. $\chi_{rad}^{\prime\prime}/\chi_{rad} = k^2$) implying\footnote{This condition follows since $k \tau_{i} \leq 1$ (or even $k \tau_{i} \ll 1$ for the typical 
scale of the gravitational collapse, as it will be shown below).} 
 $e^{ - z_{re}} = (C/D) k^2 \tau_{i}^2/\beta^2$. Therefore we will have that  $\tau_{re}/\tau_{i}$ will be given by:
\begin{equation}
\frac{\tau_{re}}{\tau_{i}} = - \frac{2}{\beta}\ln{\biggl(\frac{k}{a_{i} H_{i}}\biggr)} - \frac{1}{\beta}\ln{\biggl|\frac{C}{D}\biggr|}.
\label{tre}
\end{equation}
From Eqs. (\ref{int1cc}) and (\ref{tre}) the final expressions of the mode functions are given by: 
\begin{eqnarray}
f_{k}(\tau_{re}) &=& |k\tau_{i}|^{1/2 - \nu} \biggl[ 1 + {\mathcal O}(\beta^{-1}) \biggr] \biggl[ f_{k}(\tau_{ex}) 
\nonumber\\
&+& \frac{q}{2\nu} \frac{f^{\prime}_{k}(\tau_{ex}) - {\mathcal F}_{ex} f_{k}(\tau_{ex})}{k}\biggl( 1 - \frac{|k \tau_{i}|^{2 \nu}}{q^{2\nu}}\biggr) + {\mathcal O}(\beta^{-1}) \biggr],
\nonumber\\
 g_{k}(\tau_{re}) &=& |k\tau_{i}|^{ \nu -1/2} \biggl[ 1 + {\mathcal O}(\beta^{-1}) \biggr] \biggl[ g_{k}(\tau_{ex}) 
 \nonumber\\
&+& \frac{q}{2(1-\nu)}  \frac{g^{\prime}_{k}(\tau_{ex}) + {\mathcal F}_{ex} g_{k}(\tau_{ex})}{k}\biggl( 1 - \frac{|k \tau_{i}|^{2 (1 -\nu)}}{q^{2 ( 1- \nu)}}\biggr) + {\mathcal O}(\beta^{-1}) \biggr]. 
\end{eqnarray}
As long as $\tau \geq \tau_{re}$ 
the presence of the conductivity $\sigma$ breaks  the explicit duality symmetry so that the second equation 
of Eq. (\ref{oldeq}) will be replaced by $g_{k}' = -  k^2 f_{k} - {\mathcal F} g_{k}  - 4\pi \sigma g_{k}$. In explicit numerical integrations $\sigma$ smoothly increases  (see e.g. the last papers of Refs. \cite{duality1} and \cite{DT2}) and the equation obeyed by $f_{k}$ will then be:
\begin{equation}
\overline{f}_{k}^{\prime\prime} + \overline{f}_{k} \biggl\{ k^2 - \biggl[ \frac{\chi^{\prime\prime}}{\chi} + 4 \pi \sigma\biggl({\mathcal F} + \frac{\sigma^{\prime}}{2 \sigma} + \pi \sigma\biggr)\biggr] \biggr\} =0,
\label{cond}
\end{equation}
where $\overline{f}_{k} = \exp{[2 \pi \int^{\tau} \sigma(\tau^{\prime}) d\tau^{\prime}]} \, f_{k}(\tau)$. The structure of the turning points implied by Eq. (\ref{cond}) is different, namely,  $(k^2  - 4 \pi \sigma^2) \tau_{i}^2 = e^{- z_{re}}[ \beta^2 - 4\pi\sigma\tau_{i} \beta]/(C + e^{- z_{re}})$.
According to Eq. (\ref{cond}) the turning point is predominantly fixed by  the largeness of $\sigma\tau_{i}$ rather than by the smallness of $k \tau_{i}$: while $k\tau_{i}$ is (at most) of order $1$ (and it is much smaller than $1$ for the galactic scale) we have instead that $\sigma \tau_{i} \gg 1$, as already stressed in explicit numerical integrations of the 
of the power spectra \footnote{To compare the two scales we recall that, at a given time during radiation, $ \sigma \propto T$ whereas $H \propto T^2/M_{P}$. Since, during the postinflationary phase, $H \ll H_{i}$ we shall also have that $\sigma/H\gg 1$ and, a fortiori, $\sigma \tau_{i} \gg 1$. } (see, in particular, the last paper of \cite{DT2}). 

The hypermagnetic power spectra for generalized quantum mechanical Cauchy data can therefore be expressed as:
\begin{equation}
P_{B}(k,\tau) = {\mathcal N}_{B} H_{i}^4 a_{i}^4 \biggl(\frac{k}{k_{i}}\biggr)^{5 - 2 \nu} \biggl\{ 1 + 2 |b_{-}(k)|^2 + 2 |b_{-}(k)|\sqrt{1 + |b_{-}(k)|^2} \cos{[2 q(\nu)]}\biggr\},
\label{BPS}
\end{equation}
where ${\mathcal N}_{B}$  depends on the suddenness of the transition (parametrized, in the above example, by the value of $\beta$);  when $\beta \gg 1$, we will have that $ {\mathcal N}_{B} = (4\pi^2)^{-1} + {\mathcal O}(\beta^{-1})$. Equations (\ref{PSfirst}) and (\ref{BPS}) are clearly compatible in the case $b_{-}(k)\to 0$ and 
$b_{+}(k) \to 1$. The relation between comoving and physical power spectra (i.e. $P_{\mathrm{B}}(k,\tau) = a^4(\tau) \chi^2(\tau) P_{b}(k,\tau)$) follows from the relation between the physical and comoving field operators\footnote{This relation has nothing to do with the approximate flux conservation 
during a radiation-dominated stage of expansion, as it is erroneously stated by some. This relation stems directly from the properties of the canonical normal modes of the action (see also Eqs. (\ref{second}) and (\ref{third})).}, i.e. respectively $\vec{B}$ and $\vec{b}$ (see Eq. (\ref{second}) and discussion therein). In the spirit of the present discussion it is therefore interesting to compute the explicit relation between $\sqrt{P_{b}(k,\tau_{re})}$ and $\sqrt{P_{b}(k,\tau_{i})}$ that is given by:
\begin{equation}
\sqrt{P_{b}(k,\tau_{re})} = {\mathcal K}(\nu,\beta) \biggl[ \frac{4}{\beta^2} 
\ln^2\biggl(\frac{k}{a_{i} H_{i}}\biggr) + 9 \frac{\ln^2{\beta}}{\beta^2} - \frac{6 \ln{\beta}}{\beta^2} \ln{\biggl(\frac{k}{a_{i} H_{i}}\biggr)}\biggr] \sqrt{P_{b}(k,\tau_{i})},
\label{sol15}
\end{equation}
where ${\mathcal K}(\nu, \beta) = 2 \beta/(2\beta - 1 + 2 \nu)$ is a numerical factor of order $1$; in Eq. (\ref{sol15})
we also used that $(\tau_{re}/\tau_{i}) = - (2/\beta) \ln{k \tau_{i}} + (3/\beta) \ln{\beta}$ as it follows from Eq. (\ref{tre}) for $\beta \geq 1$.
Since the value of $k \tau_{i}= k/(a_{i} H_{i})$ is 
\begin{equation}
\frac{k}{a_{i} H_{i}} = 3.71\times 10^{-24} \, \biggl(\frac{k}{\mathrm{Mpc}^{-1}}\biggr) \, \biggl(\frac{\epsilon_{H}}{0.01}\biggr)^{-1/4} \, 
\biggl(\frac{{\mathcal A}_{\mathcal R}}{2.41\times 10^{-9}} \biggr)^{-1/4},
\label{ex0}
\end{equation}
where ${\mathcal A}_{{\mathcal R}}$ is the amplitude of the scalar power spectrum at the pivot scale 
$k_{p}= 0.002\, \mathrm{Mpc}^{-1}$;  the latter scale also defines the tensor to scalar ratio 
$r_{T} = {\mathcal A}_{T}/{\mathcal A}_{{\mathcal R}}$ (recall that $r_{T} = 16 \epsilon_{H}$ 
if the consietncy relations are enforced). According to Eq. (\ref{ex0}) $|\ln{k \tau_{i}}| = {\mathcal O}(50)$ at the scale of the protogalactic 
collapse\footnote{The natural logarithm of $x$ will be denoted hereunder by $\ln{x}$; 
common logarithms will be instead denoted by $\log{x}$.}; moreover, 
for $\beta= {\mathcal O}(10)$ (i.e. sudden
transition) the whole quantity inside the square bracket of Eq. (\ref{sol15}) is ${\mathcal O}(120)$. 
The logarithmic corrections are then insignificant: first they give, at most a contribution 
${\mathcal O}(100)$ and second they are overwhelmed by the conductivity.

Depending on the total number of efolds $N_{t}$, the initial state
can influence the late-time spectrum. In this respect the critical number of efolds is given by $N_{c}$ and it is defined as
$e^{N_{c}} = (2 \pi\, \Omega_{R 0} \,{\mathcal A}_{{\mathcal R}}\,r_{T})^{1/4}\, \sqrt{(M_{P}/H_{0})}/4$
where $\Omega_{R 0}$ is the present energy density of radiation in critical units, $H_{0}^{-1}$ is the 
Hubble radius today and $r_{T}$ is the conventional tensor to scalar ratio\footnote{ The tensor to scalar ratio itself 
is affected by the evolution of the large-scale gauge fields (see last paper of \cite{DT3}).
The latter observation implies that $r_{T}$ is bounded from below by the dominance of the 
adiabatic contribution and it cannot be smaller than $10^{-3}$, at least in the case of single-field inflationary models. 
We should therefore assume that $0.001< r_{T} < 0.1$.}.  If  $N_{t} = N_{c}$ 
the inflationary event horizon (redshifted at the present epoch) coincides with 
the Hubble radius today:
 \begin{equation}
 N_{c} = 61.49 + \frac{1}{4} \ln{\biggl(\frac{h_{0}^2 \Omega_{R 0}}{4.15 \times 10^{-5}} \biggr)} - \ln{\biggl(\frac{h_{0}}{0.7}\biggr)}
 + \frac{1}{4} \ln{\biggl(\frac{{\mathcal A}_{{\mathcal R}}}{2.41 \times 10^{-9}}\biggr)} + \frac{1}{4} \ln{\biggl(\frac{r_{T}}{0.2}\biggr)}.
  \label{eighthb}
\end{equation} 
When $N_{\mathrm{t}} > N_{c}$ the redshifted value of the inflationary event horizon 
is larger than the present value of the Hubble radius and for $N_{t} \gg N_{c}$ we plausibly expect
(at least in conventional inflationary models, that any finite portion of the Universe gradually loses the memory of an initially imposed anisotropy 
or inhomogeneity so that the Universe attains the observed regularity regardless of the initial boundary conditions).
\begin{figure}[!ht]
\centering
\includegraphics[height=7cm]{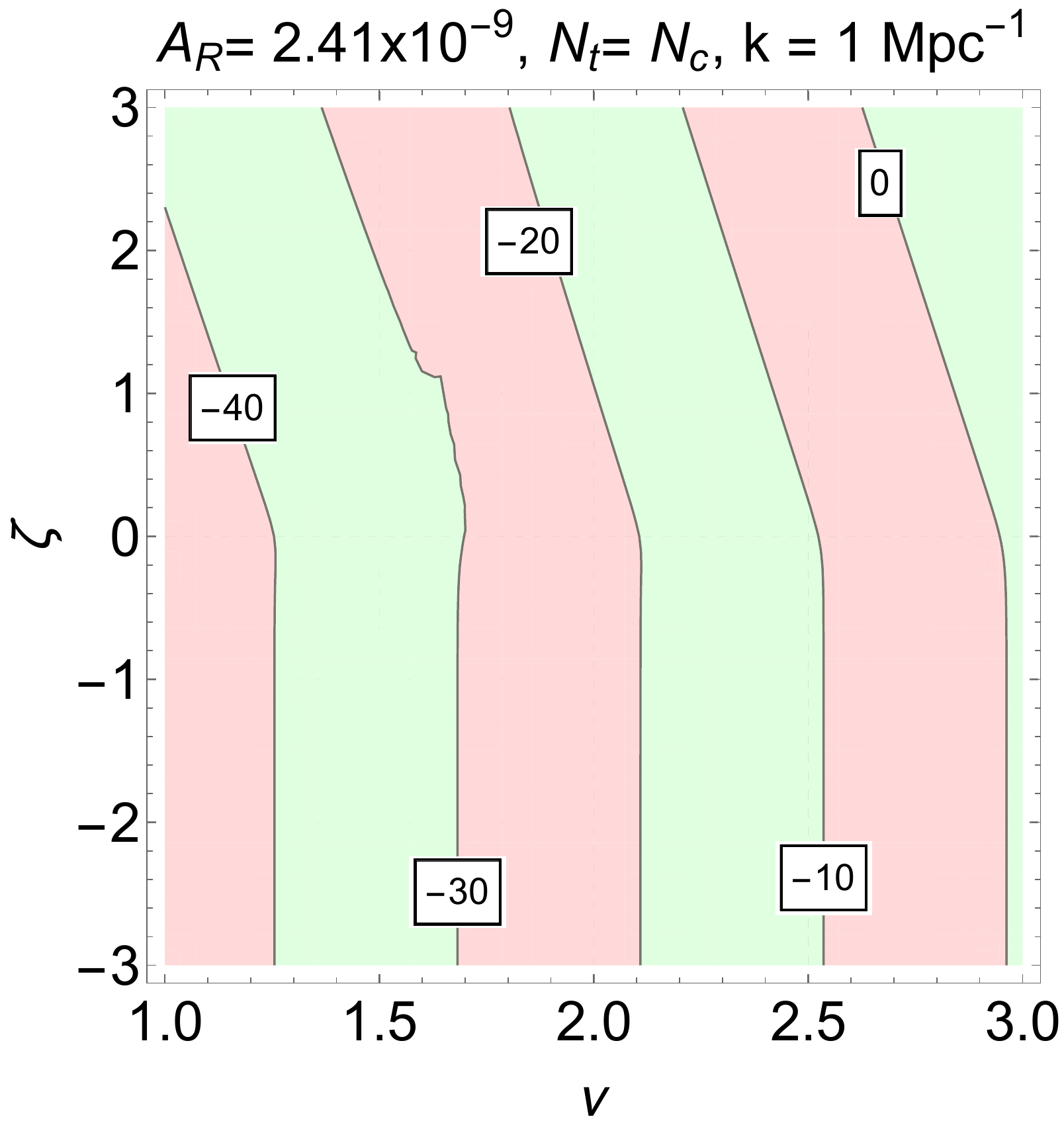}
\includegraphics[height=7cm]{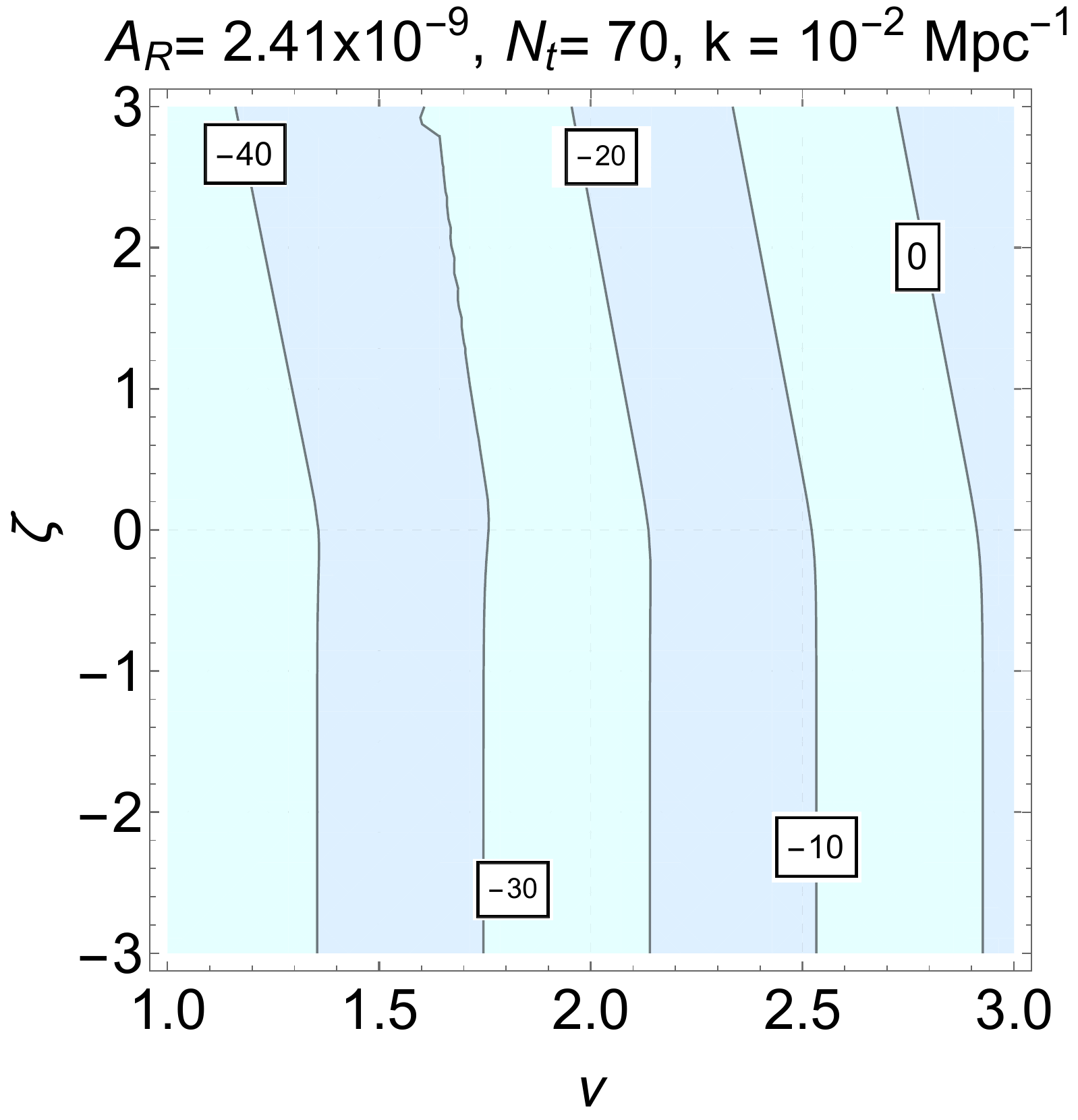}
\caption[a]{The parameter space is described in the $(\nu, \zeta)$ plane for two 
different values of the number of efolds and for two complementary comoving scales. 
The labels appearing on the various isospectral curves denote te common 
logarithm of $\sqrt{P_{B}(k,\tau_{c})}$ expressed in Gauss.}
\label{Figure1}      
\end{figure}
To investigate the role of the initial data we shall now define the protoinflationary boundary $\tau_{*}$ 
as the approximate moment at which the background starts inflating. We can then assume, without loss of generality, that 
$|b_{-}(k)|^2 = (k/k_{*})^{\zeta}$ where $k_{*} \simeq \tau_{*}^{-1}$. 
The scale $k_{*}$ is, by definition, the maximal wavenumber of the initial spectrum. 
The energy density of the initial state at $\tau_{*}$ is approximately 
$d\rho_{*} \simeq  2 |b_{-}(k)|^2 d^{3}k/(2\pi)^3$. If $\zeta > - 4$ 
the total energy density is is dominated by the largest scale 
(i.e.  $\rho_{*} = {\mathcal O}(k_{*}^4)$). 
To avoid an excessive contribution of the initial energy density to the 
dynamics of the background we must require 
$8\pi \rho_{*} \ll  3 H^2 M_{P}^2$ implying 
\begin{equation}
\frac{k}{k_{*}} = \frac{12.63}{Q_{i}}\, e^{- ( N_{c} - N)}\, \biggl(\frac{k}{\mathrm{Mpc}^{-1}} \biggr)\,
\biggl(\frac{h_{0}}{0.7}\biggr)^{-1} \,\biggl(\frac{{\mathcal A}_{{\mathcal R}}}{2.41\times 10^{-9}}\biggr)^{1/4}
\biggl(\frac{\epsilon_{H}}{0.01}\biggr)^{1/4},
\label{KT0}
\end{equation}
where $Q_{i} = H_{*} a_{*} /\sqrt{H_{i} M_{P}}< 1$ if the energetic 
content of the initial state. In practice we can choose, for instance, 
$Q_{i} = {\mathcal O}(10^{-4})$ as a reasonable 
fiducial value. Note that when the initial state is thermal 
$|b_{-}(k)|^2 \to 1/( e^{k/k_{T}}-1)$ (as implied by  
the Bose-Einstein occupation number)
where $k_{T}$ coincides, in the present units, with the putative comoving 
temperature of the initial state. 
This case will not be explicitly discussed 
here\footnote{It can be shown that the wavelengths associated with the protoinflationary 
thermal background are always larger than the typical length-scale 
related to the gravitational collapse of the protogalaxy \cite{incon}.}
but it has been carefully scrutinized in Ref. \cite{incon}.

The two-point function if the magnetic field
intensity at coincident times is derived from Eq. (\ref{cc1}) and it is
\begin{equation}
\langle \hat{B}_{i}(\tau_{c}, \vec{x}) \hat{B}_{i}( \tau_{c},\vec{x}+ \vec{r})\rangle 
= 2 \int d \ln{k} \, P_{B}(k, \tau_{c}) j_{0}(k\,r), \qquad 
j_{0}(k r) = \frac{\sin{k\,r}}{k\,r},
\label{CF}
\end{equation}
implying\footnote{Equations (\ref{rel2}) 
and (\ref{cc1}) imply that $\hat{B}_{i}(\tau,\vec{x})$  has dimensions $L^{-2}$;
 thus, dimensionally,  $[\hat{B}_{i}(\tau,\vec{k})] = L$ and $[P_{B}(\tau, k)] = L^{-4}$ 
 taking into account the dimensions 
of the  the three-dimensional Dirac delta function in Eq. (\ref{cc1}).}  
that $P_{B}(k, \tau_{c})$ and  $\sqrt{P_{B}(k, \tau_{c})}$ measure, respectively, the 
energy density the field intensity over the typical wavenumber $1/r$.
For the applications to magnetogenesis problems \cite{DT4,rev} it is useful to compute 
the physical power spectrum expressed in units of Gauss (G, in what follows)
and at the time of the gravitational collapse of the protogalaxy. The result of this calculation is
\begin{equation}
\frac{\sqrt{P_{b}(k,\tau_{c})}}{\mathrm{G}} = 10^{-10.84}\, 
\biggl(\frac{{\mathcal A}_{{\mathcal R}}}{2.41 \times 10^{-9}}\biggr)^{1/2}\,
\biggl(\frac{\Omega_{\mathrm{R}0}}{4.15 \times 10^{-5}}\biggr)^{1/2}\,\cos{\theta_{W}}
 {\mathcal M}(\nu, \zeta, k,  k_{*}),
\label{sol13}
\end{equation}
where, as already mentioned, $\sqrt{P_{B}(k,\tau)} = a^{2}(\tau) \chi(\tau)\sqrt{P_{b}(k,\tau)}$ 
for $\tau < \tau_{re}$ and $\sqrt{P_{B}(k,\tau)} = a^{2}(\tau) \sqrt{P_{b}(k,\tau)}$ for $\tau> \tau_{re}$; 
the function $ {\mathcal M}(\nu, \zeta, k,  k_{*})$ appearing in Eq. (\ref{sol13})  is given by
\begin{equation}
{\mathcal M}(\nu, \zeta, k,  k_{*}) = \biggl(\frac{k}{a_{i} H_{i}}\biggr)^{5/2 -  \nu}
\biggl\{ 1 + 2 \biggl(\frac{k}{k_{*}}\biggr)^{\zeta} + 2 \biggl(\frac{k}{k_{*}}\biggr)^{\zeta/2} 
\sqrt{1 + \biggl(\frac{k}{k_{*}}\biggr)^{\zeta} } \cos{[2 q(\nu)]}\biggr\}^{1/2}.
\label{sol14}
\end{equation}
In Eq. (\ref{sol13}) we used that the non-screened vector modes of the hypercharge 
field the project on the electromagnetic fields through the cosine of the Weinberg angle $\cos{\theta_{W}}$. 

Equations (\ref{sol13}) and (\ref{sol14}) shall now be analyzed in the ($\nu$, $\zeta$) 
plane illustrated in Fig. \ref{Figure1} where we plot the isospectral lines (i.e. the lines of the parameter space 
over which the magnetic power spectrum 
is approximately constant) for few physically meaningful choices of the parameters.
When $\zeta\to 0$ the scale-invariant limit of the magnetic power spectrum corresponds to
$\nu \to 5/2= 2.5$ with typical amplitude ${\mathcal O}(10^{-10})$G: this limit can be verified
in both plots of Fig. \ref{Figure1}. When $\zeta \neq 0$, quasi-flat power spectra can be obtained as long as $\nu$ remains in the region 
where $\log{[\sqrt{P_{B}(k,\tau_{c})}]} = {\mathcal O}( -10)$. In the left plot of Fig. \ref{Figure1} we illustrated 
the benchmark scale of the gravitational collapse of the protogalaxy (i.e. $k = {\mathcal O}(1) \, \mathrm{Mpc}^{-1}$) when the number of 
efolds is just critical (i.e. $N_{t}= N_{c}$); in the right plot we instead considered a larger typical length-scale (i.e. smaller wavenumber 
 $k = {\mathcal O}(10^{2}) \, \mathrm{Mpc}^{-1}$) and a larger total number of efolds (i.e. $N_{t} =70 > N_{c}$).
What matters is the inclination of the (almost straight) isospectral lines  for $\zeta >0$. Figure \ref{Figure1} shows that 
for $N_{t} > N_{c}$ the inclination diminishes and the lines become more and more vertical when $N_{t} \gg N_{c}$. 
The latter observation shows that as long as $N_{t} = {\mathcal O}(N_{c})$ the parameter space of inflationary magnetogenesis 
is comparatively larger than in the case $\zeta =0$: we pass from a point (i.e. $\zeta =0$ and $\nu\simeq 2.5$) 
to a whole isospectral line in the $(\nu, \zeta)$ plane.

If $\sqrt{P_{b}(k_{c}, \tau_{c})}$ is approximately larger than about $10^{-24}$ G 
(but still much smaller than  ${\mathcal O}(10^{-10})$ G)  the observed galactic 
field intensity can only be reached if the fields are amplified (for $\tau > \tau_{re}$)
by the combined action of the gravitational collapse and of the galactic rotation. 
The latter  effect may hopefully transform, under various  conditions, the kinetic 
energy of the plasma into magnetic energy \cite{rev}. The most optimistic estimates 
for the required initial conditions are derived by assuming that every rotation 
of the galaxy would increase the magnetic field of one efold. The number of galactic 
rotations since the collapse of the protogalaxy can be between $30$ and $35$, 
leading approximately to a purported growth of 13 orders of magnitude. 
If the dynamo action is totally absent, the required field should be ${\mathcal O}(10^{-11})$ G. 
In this case, during the collapse of the protogalaxy, the magnetic field will increase by 
about $5$ orders of magnitude. In the literature it is sometimes practical 
to refer to some hypothetical seed field supposedly present 
at the time of the collapse of the protogalaxy. 
By definition ${\mathcal B}_{seed}= \sqrt{P_{b}(k_{c}, \tau_{c})}$
where following the standard conventions 
\cite{rev} we took $k_{c} = {\mathcal O}(1)\, \mathrm{Mpc}^{-1}$. From the above 
considerations we have therefore that 
\begin{equation}
{\mathcal O}(10^{-24}) \, \mathrm{G} \leq {\mathcal B}_{seed}(\tau_{c}) \leq {\mathcal O}(10^{-12}) \mathrm{G}.
\label{seed1}
\end{equation}
After the gauge modes reenter the effective horizon the approximate flux conservation implies  
 ${\mathcal B}_{seed}(\tau_{c}) = {\mathcal B}_{seed}(\tau_{re}) (H_{0}/M_{P}) \sqrt{\Omega_{R0}/[\pi \epsilon_{H} {\mathcal A}_{{\mathcal R}}]}$ and Eq. (\ref{seed1}), at $\tau_{re}$ can be written, up to the insignificant logarithmic corrections discussed above, as:
\begin{equation}
{\mathcal O}(10^{33}) \, \mathrm{G} \leq {\mathcal B}_{seed}(\tau_{re}) \leq {\mathcal O}(10^{45}) \mathrm{G}.
\label{seed2}
\end{equation}
Equations (\ref{seed1}) and (\ref{seed2}) are insensitive to the properties 
of the initial state but they depend 
on the postinflationary thermal history, as 
already discussed in the past \cite{incon}.

In summary the junction conditions for the gauge fields are compatible with sudden and 
delayed transitions of the effective horizon. The dynamical evolution of the gauge modes 
has been rephrased in terms of a pair of integral equations related by duality transformations. 
After showing how the continuity of the susceptibility and of its first derivative determines 
the hypermagnetic and hyperelectric power spectra, explicit examples of smooth transitions 
have been proposed to corroborate the analytic discussion.  
The general arguments based on the continuity of the effective horizon are valid up to 
logarithmic corrections which are numerically not significant when the gauge modes reenter the 
effective horizon. Moreover, after reentry these corrections are anyway overwhelmed by the dominance 
of the conductivity.  As long as the total duration of the inflationary phase 
is nearly minimal the spectral slopes may be directly affected by the properties of the initial state. In the latter case case 
the parameter space of quasi-flat spectra gets larger. Conversely, when the number 
of efolds increases beyond a certain critical value,  the present findings reproduce the previous results
since the effects of the initial state are exponentially suppressed.

\end{document}